\documentclass[prd,superscriptaddress,amsfonts,amssymb,amsmath,showpacs,onecolumn]{revtex4-2}
\usepackage{bm}
\usepackage{amsfonts}
\usepackage{latexsym}
\usepackage[latin1]{inputenc}
\usepackage{graphicx}
\usepackage{amsmath}
\usepackage{palatino}
\usepackage{mathpazo}
\usepackage{textcomp}
\usepackage{float}
\usepackage{booktabs}
\usepackage{dcolumn}
\usepackage{multirow}
\usepackage{ragged2e}
\usepackage{hyperref}
\hypersetup{colorlinks,citecolor=blue}
\usepackage{amsmath}
\usepackage{xcolor}
\usepackage{orcidlink}
\usepackage[caption=false]{subfig}
\usepackage{commath}
\def\jnl@style{\it}
\def\aaref@jnl#1{{\jnl@style#1}}

\def\aaref@jnl#1{{\jnl@style#1}}

\def\aj{\aaref@jnl{AJ}}                   
\def\apj{\aaref@jnl{ApJ}}                 
\def\apjl{\aaref@jnl{ApJ}}                
\def\apjs{\aaref@jnl{ApJS}}               
\def\apss{\aaref@jnl{Ap\&SS}}             
\def\aap{\aaref@jnl{A\&A}}                
\def\aapr{\aaref@jnl{A\&A~Rev.}}          
\def\aaps{\aaref@jnl{A\&AS}}              
\def\mnras{\aaref@jnl{Mon.~Not.~Roy.~Astron.~Soc.}}             
\def\prd{\aaref@jnl{Phys.~Rev.~D}}        
\def\prc{\aaref@jnl{Phys.~Rev.~C}}  
\def\prl{\aaref@jnl{Phys.~Rev.~Lett.}}    
\def\qjras{\aaref@jnl{QJRAS}}             
\def\skytel{\aaref@jnl{S\&T}}             
\def\ssr{\aaref@jnl{Space~Sci.~Rev.}}     
\def\zap{\aaref@jnl{ZAp}}                 
\def\nat{\aaref@jnl{Nature}}              
\def\aplett{\aaref@jnl{Astrophys.~Lett.}} 
\def\apspr{\aaref@jnl{Astrophys.~Space~Phys.~Res.}} 
\def\physrep{\aaref@jnl{Phys.~Rep.}}      
\def\physscr{\aaref@jnl{Phys.~Scr}}       
\def\commat{\aaref@jnl{Comm.~Math.~Phys.}}              
\def\science{\aaref@jnl{Science}}               
\def\cqg{\aaref@jnl{Classical Quant.~Grav.}}            
\def\jpcs{\aaref@jnl{JPCS}}                                     
\def\ijmpd{\aaref@jnl{Int.~J.~Mod.~Phys.~D}}                    
\def\grg{\aaref@jnl{Gen.~Relat.~Gravit.}}               
\def\rpp{\aaref@jnl{Rep.~Prog.~Phys.}}          
\def\npa{\aaref@jnl{Nucl.~Phys.~A}}        
\def\lrr{\aaref@jnl{Living Rev.~Rel.}}                   
\def\jcap{\aaref@jnl{J.~Cosmology Astropart.~Phys.}}    
\def\rmp{\aaref@jnl{Rev.~Mod.~Phys.}}   
\def\epjc{\aaref@jnl{Eur.~Phys.~J.~C}}

\allowdisplaybreaks[1]

\begin{document}

\color{black}       

\title{Phase-space analysis of the viscous fluid cosmological models in the coincident $f(Q)$ gravity}

\author{Dheeraj Singh Rana\orcidlink{0000-0002-4401-8814}}
\email{drjrana2@gmail.com}
\affiliation{Department of Mathematics, Birla Institute of Technology and
Science-Pilani,\\ Hyderabad Campus, Hyderabad-500078, India.}
\author{Raja Solanki\orcidlink{0000-0001-8849-7688}}
\email{rajasolanki8268@gmail.com}
\affiliation{Department of Mathematics, Birla Institute of Technology and
Science-Pilani,\\ Hyderabad Campus, Hyderabad-500078, India.}
\author{P.K. Sahoo\orcidlink{0000-0003-2130-8832}}
\email{pksahoo@hyderabad.bits-pilani.ac.in}
\affiliation{Department of Mathematics, Birla Institute of Technology and
Science-Pilani,\\ Hyderabad Campus, Hyderabad-500078, India.}


\date{\today}
\begin{abstract}

 In this article, we consider a newly proposed parameterization of the viscosity coefficient $\zeta$, specifically $\zeta=\bar{\zeta}_0 {\Omega^s_m} H $, where $\bar{\zeta}_0 = \frac{\zeta_0}{{\Omega^s_{m_0}}} $ within the coincident $f(Q)$ gravity formalism. We consider a non-linear function $f(Q)= -Q +\alpha Q^n$, where $\alpha$ and $n$ are arbitrary model parameters, which is a power-law correction to the STEGR scenario. We find an autonomous system by invoking the dimensionless density parameters as the governing phase-space variables. We discuss the physical significance of the model corresponding to the parameter choices $n=-1$ and $n=2$ along with the exponent choices $s=0, 0.5$, and $1.05$. We find that model I shows the stable de-Sitter type or stable phantom type (depending on the choice of exponent $s$) behavior with no transition epoch, whereas model II shows the evolutionary phase from the radiation epoch to the accelerated de-Sitter epoch via passing through the matter-dominated epoch. Hence, we conclude that model I provides a good description of the late-time cosmology but fails to describe the transition epoch, whereas model II modifies the description in the context of the early universe and provides a good description of the matter and radiation era along with the transition phase. 

\end{abstract}

\maketitle


\section{Introduction}\label{sec1}
\justifying

Modified gravity plays a prominent role in enhancing our comprehension of the universe's evolution, encompassing its early phase as well as later stages marked by accelerated expansion, often associated with dark energy \cite{CANT}. Moreover, it also serves as a potential resolution for the observational conflicts \cite{COSI}. These theories involve modifying general relativity to introduce additional degrees of freedom, potentially offering corrections at both the background and perturbation levels. Various methods exist to construct such modifications of general relativity. One such gravitational modification is obtained by utilizing the non-metricity to produce an equivalent formulation of the general relativity, widely recognized as symmetric teleparallel gravity \cite{NEST}. In this approach, a generic affine connection is utilized, characterized by both vanishing torsion and vanishing curvature in relation to the Levi-Civita connection. This involves relaxing the metricity condition corresponding to the generic affine connection. Recently, this formulation has been further modified to give $f(Q)$ gravity \cite{JIM-1}. The extended symmetric teleparallel formalism has attracted interest within the cosmology community as a promising pathway to investigate novel physics beyond the established $\Lambda$CDM cosmology. Specific $f(Q)$ functional forms have been shown to resolve the $\sigma8$ tension \cite{BARR}, while other particular forms facilitate a more precise description of the cosmological observational data \cite{ANAG,NUNES}. Recently, notable cosmological implications of the $f(Q)$ gravity in various contexts have appeared, for instance, Neutrino physics \cite{NEOM}, Black hole physics \cite{RODR,LAVI-1}, Astrophysical objects \cite{SNEHA,ZINNAT}, Quantum cosmology \cite{CAPE-1,PALIA-1}, Cosmological perturbations \cite{ET}, BBN constraints \cite{ANAG-2}, Phantom cosmology \cite{ANDER}, Inflation \cite{CAPE-2}, and many others \cite{HOH,WOM-2,DE-1,DE-2,PALIA-2}. \\

In previous investigations of the early inflationary period, researchers explored this phenomenon without introducing the concept of dark energy. To describe the early universe phenomenon, they included viscosity in the cosmic fluid to take dynamics into consideration. Eckart made a contribution to this topic by developing a non-causal theory of viscosity in which he focused on first-order deviations from the equilibrium state \cite{C.E.}. Subsequently, Israel and Stewart further advanced the understanding of viscosity in this context by presenting a causal theory, which involved considering second-order deviations from the equilibrium state \cite{W.I.,W.I.-2,W.I.-3}. Prominent investigations into the observed late-time accelerating expansion of the universe have been carried out using the causal theory. When we consider the second-order deviations from the equilibrium state, two distinct viscosity coefficients come into account: shear viscosity and bulk viscosity. These coefficients play essential roles in describing the behavior of cosmic fluids in various scenarios. In a homogeneous background spacetime, it's important to note that the velocity gradient associated with shear viscosity disappears. Hence, in a homogeneous and isotropic FLRW background, only the bulk viscosity becomes relevant and plays a significant role in describing the behavior of the cosmic fluid. As the cosmic matter content moves along with the expanding universe, the pressure essential for re-establishing thermal equilibrium can be recognized as the bulk viscosity. Thus, the modification of general relativity provides a justification for the expansion of the universe, whereas the viscosity coefficients play a significant role in the pressure term that propels the cosmic acceleration. Numerous studies have concentrated their efforts on investigating the influence of bulk viscosity on the evolution of the universe, as shown in several references \cite{IB-1,IB-2,IB-3,IB-4,IB-5,JM,AVS,MAT}.

It is well known that incorporating auxiliary variables can transform the cosmological set of field equations into a set of autonomous differential equations \cite{COPE}. This leads to a system $X' = f(X)$, where $X$ stands for the column vector encompassing auxiliary variables, and $f(X)$ represents the vector field. The examination of stability for the given autonomous system involves a multi-step process. First, the equilibrium points (or critical points) denoted by $X_c$ are estimated using the equation $X' = 0$. Following this, we assume linear perturbations near the critical point $X_c$, representing them as $ X = X_c + P $, where $P$ is the column vector representing perturbed auxiliary variables. Consequently, it is possible to establish the matrix equation $P = AP$ (up to first order), where A represents the matrix comprising the coefficients of the perturbed equations. One can examine the stability properties of each hyperbolic-type equilibrium point by utilizing the eigenvalues of the coefficient matrix A. An equilibrium point $X_c$ of the given autonomous system is stable (unstable) or a saddle based on whether the real parts of the associated eigenvalues are negative (positive) or possess real parts with opposite signs. Various noteworthy findings within the realm of modified gravity using the dynamical system approach have appeared in references \cite{DE-3,WOM,Mishra-2,HAMID,APLL}. The present manuscript is organized as follows. The mathematical framework of $f(Q)$ gravity in a flat symmetric teleparallelism background has been presented in the section \ref{sec2}. In section \ref{sec3}, we present the governing equations of motion corresponding to flat homogeneous and isotropic FLRW background. In section \ref{sec4}, we set up an autonomous system corresponding to a generic power-law $f(Q)$ function along with a newly proposed parameterization of the viscosity coefficient. Further, we present our analysis for two toy models with corresponding phase diagrams and relevant cosmological parameters. Finally, in section \ref{sec5}, we discuss our findings.

\section{$f(Q)$ gravity formalism}\label{sec2}
\justifying

It is well known that a connection plays a fundamental role in facilitating the transportation of tensors throughout the spacetime manifold. In the context of general relativity, the Levi-Civita connection, which is a symmetric connection, governs gravitational interactions. However, a generic affine connection involves the presence of two more distinct components, specifically an antisymmetric tensor and another one that exhibits non-metricity. This generic connection can be mathematically represented as follows \cite{Jimenez3}:

\begin{equation}\label{2a}
\Upsilon^\alpha_{\ \mu\nu}=\Gamma^\alpha_{\ \mu\nu}+K^\alpha_{\ \mu\nu}+L^\alpha_{\ \mu\nu},
\end{equation}
Here $\Gamma^\alpha_{\ \mu\nu}$ represents the Levi-Civita connection, which exhibits the metricity condition, and that is given by following relation, 
\begin{equation}\label{2b}
\Gamma^\alpha_{\ \mu\nu}\equiv\frac{1}{2}g^{\alpha\lambda}(g_{\mu\lambda,\nu}+g_{\lambda\nu,\mu}-g_{\mu\nu,\lambda})
\end{equation}
The second term refers to an antisymmetric contortion tensor, and can be obtained by utilizing the torsion tensor $ T^\alpha_{\ \mu\nu}\equiv \Upsilon^\alpha_{\ \mu\nu}-\Upsilon^\alpha_{\ \nu\mu}$ as follows,
\begin{equation}\label{2c}
K^\alpha_{\ \mu\nu}\equiv\frac{1}{2}(T^{\alpha}_{\ \mu\nu}+T_{\mu \ \nu}^{\ \alpha}+T_{\nu \ \mu}^{\ \alpha}),
\end{equation}
Moreover, the distortion tensor can be expressed in terms of non-metricity tensor as,
\begin{equation}\label{2d}
L^\alpha_{\ \mu\nu}\equiv\frac{1}{2}(Q^{\alpha}_{\ \mu\nu}-Q_{\mu \ \nu}^{\ \alpha}-Q_{\nu \ \mu}^{\ \alpha})
\end{equation}	
where, 
\begin{equation}\label{2e}
Q_{\alpha\mu\nu}\equiv\nabla_\alpha g_{\mu\nu} = \partial_\alpha g_{\mu\nu}-\Upsilon^\beta_{\,\,\,\alpha \mu}g_{\beta \nu}-\Upsilon^\beta_{\,\,\,\alpha \nu}g_{\mu \beta}
\end{equation} 
We write the following notion of the superpotential tensor as,
\begin{equation}\label{2f}
4P^\lambda\:_{\mu\nu} = -Q^\lambda\:_{\mu\nu} + 2Q_{(\mu}\:^\lambda\:_{\nu)} + (Q^\lambda - \tilde{Q}^\lambda) g_{\mu\nu} - \delta^\lambda_{(\mu}Q_{\nu)}.
\end{equation}
where $Q_\alpha = Q_\alpha\:^\mu\:_\mu $ and $ \tilde{Q}_\alpha = Q^\mu\:_{\alpha\mu} $ are two vectors obtained by contracting  non-metricity tensor. One can express the non-metricity scalar in a following manner, by contracting the superpotential and the non-metricity tensor, 
\begin{equation}\label{2g}
Q = -Q_{\lambda\mu\nu}P^{\lambda\mu\nu}. 
\end{equation}
Further, the curvature tensor can be acquired as
\begin{equation}\label{2h}
R^\alpha_{\: \beta\mu\nu} = 2\partial_{[\mu} \Upsilon^\alpha_{\: \nu]\beta} + 2\Upsilon^\alpha_{\: [\mu \mid \lambda \mid}\Upsilon^\lambda_{\nu]\beta}
\end{equation} 
Now by using the equation \eqref{2h} and the affine connection \eqref{2a}, we have
\begin{equation}\label{2i}
R^\alpha_{\: \beta\mu\nu} = \mathring{R}^\alpha_{\: \beta\mu\nu} + \mathring{\nabla}_\mu X^\alpha_{\: \nu \beta} - \mathring{\nabla}_\nu X^\alpha_{\: \mu \beta} + X^\alpha_{\: \mu\rho} X^\rho_{\: \nu\beta} - X^\alpha_{\: \nu \rho} X^\rho_{\: \mu\beta}
\end{equation}
where $X^\alpha_{\ \mu\nu}=K^\alpha_{\ \mu\nu}+L^\alpha_{\ \mu\nu}$. Note that the quantities $\mathring{R}^\alpha_{\: \beta\mu\nu}$ and $\mathring{\nabla}$ uses the Levi-Civita connection \eqref{2b}. By utilizing the appropriate contractions and the constraint $ T^\alpha_{\ \mu\nu}=0$ to the curvature term \eqref{2i}, we obtained
\begin{equation}\label{2j}
R=\mathring{R}-Q + \mathring{\nabla}_\alpha \left(Q^\alpha-\tilde{Q}^\alpha \right)   
\end{equation}
Again, on employing teleparallel constraint $R=0$, the relation \eqref{2j} becomes
\begin{equation}\label{2k}
\mathring{R}=Q - \mathring{\nabla}_\alpha \left(Q^\alpha-\tilde{Q}^\alpha \right)   
\end{equation}
The relationship \eqref{2k} represents that the non-metricity scalar and the Ricci scalar differs by a boundary term. In view of the equation \eqref{2k}, one can conclude that the gravitational theory consists only of the non-metricity scalar $Q$, differs from Hilbert's action of the GR, due to the presence of a boundary term. This suggests that STEGR offers a formulation that is equivalent to GR, and as a result, the theory is referred to as symmetric teleparallel equivalent to GR \cite{KUHN}.
Now, we write the action for $f(Q)$ gravity, which serves as a generalization of the STEGR theory, under the framework of flat and symmetric connection,
\begin{equation}\label{2l}
\mathcal{S}=\int\frac{1}{2}\,f(Q)\sqrt{-g}\,d^4x+\int \mathcal{L}_m\,\sqrt{-g}\,d^4x\, ,
\end{equation}
In this context, the notation $f(Q)$ represents a function that depends on the scalar quantity $Q$, while $L_m$ stands for the Lagrangian density. The variation of the action \eqref{2l} with respect to the metric yields the following metric field equation,
\begin{equation}\label{2m}
\frac{2}{\sqrt{-g}}\nabla_\lambda (\sqrt{-g}f_Q P^\lambda\:_{\mu\nu}) + \frac{1}{2}g_{\mu\nu}f+f_Q(P_{\mu\lambda\beta}Q_\nu\:^{\lambda\beta} - 2Q_{\lambda\beta\mu}P^{\lambda\beta}\:_\nu) = -T_{\mu\nu}
\end{equation}
where $f_Q=\frac{df}{dQ}$ and $\mathcal{T}_{\mu \nu}$ is the stress-energy tensor given by,
\begin{equation}\label{2n}
\mathcal{T}_{\mu\nu} = \frac{-2}{\sqrt{-g}} \frac{\delta(\sqrt{-g}L_m)}{\delta g^{\mu\nu}}
\end{equation}
Furthermore, in the absence of hypermomentum, the variation of \eqref{2l} with respect to the connection yields the following equation,
\begin{equation}\label{2o}
\nabla_\mu \nabla_\nu (\sqrt{-g}f_Q P^{\mu\nu}\:_\lambda) =  0 
\end{equation}

\section{Equations of Motion}\label{sec3}
\justifying
In order to investigate the cosmological consequences under the cosmological principle assumption, we start with the following flat FLRW metric,
\begin{equation}\label{3a}
ds^2= -dt^2 + a^2(t)[dx^2+dy^2+dz^2]    
\end{equation}
Here scale factor $a(t)$ quantifies the universe's expansion. To initiate this analysis, we begin with the teleparallel constraint associated with a flat geometry, which represents a purely inertial connection. Subsequently, a gauge transformation, parameterized by $\Lambda^\alpha_\mu$ \cite{Jimenez2}, can be performed,
\begin{equation}\label{3b}
 \Upsilon^\alpha_{\: \mu \nu}  = (\Lambda^{-1})^\alpha_{\:\: \beta} \partial_{[ \mu}\Lambda^\beta_{\: \: \nu ]}
\end{equation}
As a result, the generic affine connection can be represented in the following way, 
\begin{equation}\label{3c}
\Upsilon^\alpha_{\: \mu \nu} = \frac{\partial x^\alpha}{\partial \zeta^\rho} \partial_\mu \partial_\nu \zeta^\rho
\end{equation}
This representation takes advantage of an arbitrary element of the group $ GL(4,\mathbb{R}) $, which is defined by the transformation $ \Lambda^\alpha_{\: \: \mu}=\partial_\mu \zeta^\alpha$. It is essential to note that $ \zeta^\alpha $ represents an arbitrary vector field in this context. Also, this suggests that it is feasible to eliminate the affine connection by employing a coordinate transformation. Such a coordinate transformation is well-known as the gauge coincident. Assuming the coincident gauge in the present formalism, the non-metricity scalar associated with the metric \eqref{3a} simplifies to $Q=6H^2$.\\

In the context of the dark matter fluid, the concept of bulk viscous pressure is proposed as a means to incorporate an effective pressure, enabling a broader range of phenomenological outcomes within a cosmological framework that extends beyond the conventional perfect fluid model. This effective pressure is defined as $\bar{p}=p-3\zeta H$, where $\zeta$ represents the bulk viscosity coefficient, ensuring compliance with the second law of thermodynamics under the condition that $\zeta > 0$. In this manuscript, we consider the newly proposed parameterization of the $\zeta$ as follows \cite{NVV},
\begin{equation}\label{3d}
\zeta=\zeta_0 H^{1-2s} H_0^{2s} \left(\frac{\rho_m}{\rho_{m_0}}\right)^s=\Bar{\zeta}_0 {\Omega^s_m}H 
\end{equation}
where $\bar{\zeta}_0 = \frac{\zeta_0}{{\Omega^s_{m_0}}} $. This parameterization offers some benefits, one notable advantage is that it encompass widely recognized models, $ \zeta=\zeta(H)$ for the case $ s=0$ and $ \zeta \sim {{\rho}_m}^{\frac{1}{2}} $ for the case $s = \frac{1}{2}$. The corresponding stress-energy tensor becomes
\begin{equation}\label{3e}
T_{\mu\nu}=(\rho+\bar{p})u_\mu u_\nu + \bar{p} g_{\mu\nu}
\end{equation}
Here $\bar{p}$ stands for the effective pressure with the effect of viscosity coefficient $\zeta$, and defined as $\bar{p}=p-3\zeta H$. Additionally, the four-velocity vector is denoted as $u^\mu=(1,0,0,0)$.
The Friedmann like equations for the functional form $f(Q) = - Q + \Psi(Q)$ with respect to the metric \eqref{3a} can be given as
\begin{equation}\label{3f}
\Psi+Q-2Q\Psi_Q=2\rho
\end{equation}
\begin{equation}\label{3g}
\dot{H}=\frac{p+\rho}{2(-1+\Psi_Q+2\Psi_{QQ})}
\end{equation}
The Friedmann equations \eqref{3f} and \eqref{3g} can be recognized as STEGR cosmology with an additional dark energy fluid part. The evolution of the dark energy components, emerging due to non-metricity, is defined by,
\begin{equation}\label{3h}
\rho_{de}=-\frac{\Psi}{2}+ Q\Psi_Q
\end{equation}
\begin{equation}\label{3i}
p_{de}=-\rho_{de}-2\dot{H} \left( \Psi_Q+2Q\Psi_{QQ} \right)
\end{equation}
where $\rho_{de}$ and $p_{de}$ represents the corresponding energy density and pressure term. The effective Friedmann equations possessing radiation and viscous type matter along with a contribution of geometrical dark energy component reads as,
\begin{equation}\label{3j}
3H^2= \rho_m + \rho_r + \rho_{de}
\end{equation}
\begin{equation}\label{3k}
\dot{H}=-\frac{1}{2} [\rho_m + \rho_r + \rho_{de} + \bar{p}_m + p_r+p_{de}]
\end{equation}
Moreover, the standard continuity equation can be obtained as,
\begin{equation}\label{3l}
\dot{\rho}_m + 3H\left(\rho_m- 3 \zeta H \right)=0
\end{equation}
\begin{equation}\label{3m}
\dot{\rho}_r + 4H \rho_r=0
\end{equation}
\begin{equation}\label{3n}
\dot{\rho}_{de} + 3H\left(\rho_{de}+p_{de}\right)=0
\end{equation}

\section{The Cosmological Model and Phase-space Analysis}\label{sec4}
\justifying

In this section, we begin with the following dimensionless variables that encompass the entire evolution of the system's phase space. This variables allows us to transform the system's dynamics into the structure of an autonomous system, which makes it easier to understand the behavior of the system. The considered dimensionless variables are as follow,
\begin{equation}\label{4a}
x=\Omega_m=\frac{\rho_m}{3H^2}, \:\: y=\Omega_r=\frac{\rho_r}{3H^2}, \:\: \text{and} \:\: z=\Omega_{de}=\frac{\rho_{de}}{3H^2}
\end{equation}
Using the equation \eqref{4a}, the expression \eqref{3j} becomes,
\begin{equation}\label{4b}
x+y+z=1 
\end{equation}
Hence it follows that $z=1-x-y$ as well as $0 \leq x,y,z \leq 1$. Again, by using the equations \eqref{3j} and \eqref{3k} with the continuity equations, we obtain
\begin{equation}\label{4c}
\frac{\dot{H}}{H^2}=\frac{-3x\left(1+\bar{\zeta}_0 x^{s-1}\right)-2y}{2(\Psi_Q+2Q \Psi_{QQ}-1)}
\end{equation}
\justify We consider a power-law function of the non-metricity, specifically, $\Psi(Q)=\alpha Q^n$ i.e. $f(Q)=-Q+\alpha Q^n$, where $\alpha$ and $n$ are arbitrary model parameters. Also, the assumed choice of the $f(Q)$ non-metricity function is suitable to close the governing dynamical system. Corresponding to this choice, the expression \eqref{4c} becomes
\begin{equation}\label{4d}
\frac{\dot{H}}{H^2}=\frac{-3x\left(1+\bar{\zeta}_0 x^{s-1}\right)-2y}{2\left[n(1-x-y)-1\right]}
\end{equation}
The corresponding expressions for the deceleration and the effective equation of state parameter reads as,
\begin{equation}\label{4e}
q=-1+ \frac{3x\left(1+\bar{\zeta}_0 x^{s-1}\right)+2y}{2\left[n(1-x-y)-1\right]}
\end{equation}
\begin{equation}\label{4f}
\omega_{total}=-1+\frac{3x\left(1+\bar{\zeta}_0 x^{s-1}\right)+2y}{3\left[n(1-x-y)-1\right]}
\end{equation}
We acquired the following autonomous system of equations, with respect to e-folding time $N=ln(a)$, for the assumed cosmological settings consisting of power-law $f(Q)$ function with the cosmic fluid incorporates radiation and viscous matter, 
\begin{equation}\label{4g}
x'=x\left[3\bar{\zeta}_0 x^{s-1}-3+\frac{\left[3x\left(1+\bar{\zeta}_0 x^{s-1}\right)+2y\right]}{\left[n(1-x-y)-1\right]}\right],
\end{equation}
\begin{equation}\label{4h}
y'=\frac{y}{\left[n(1-x-y)-1\right]}\left[3x\left(1+\bar{\zeta}_0 x^{s-1}\right)-4n(1-x-y)+2y+4\right].
\end{equation}
\justify From the presented set of equations it is evident that the one cannot compute the equilibrium points of the system explicitly for the arbitrary choice of parameters. Hence, for the further investigation, we will explore two specific $f(Q)$ model for some well known choice of the exponent $s$, as follows: \\

\begin{figure}[H]
{\includegraphics[scale=0.41]{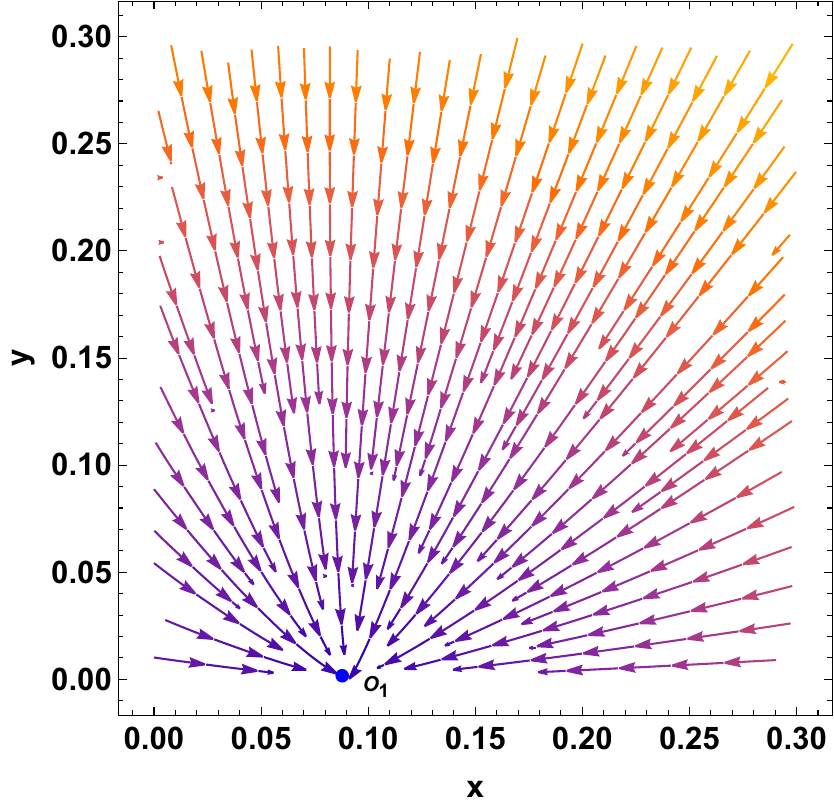}}
{\includegraphics[scale=0.41]{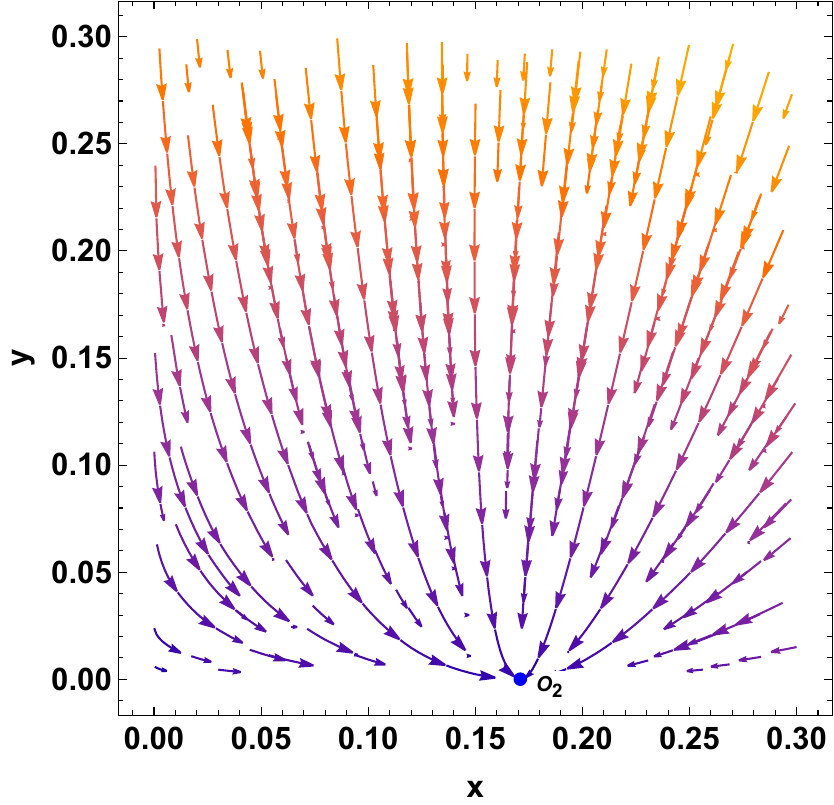}}
{\includegraphics[scale=0.41]{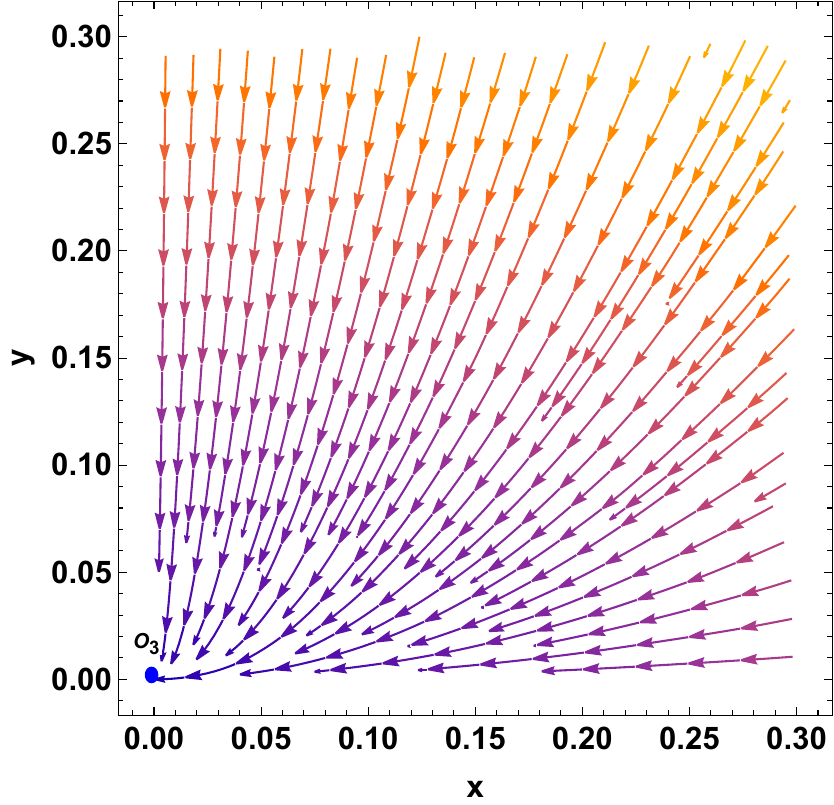}}
\caption{Phase-space diagrams representing the behavior of trajectories for the model I corresponding to cases $(s,\bar{\zeta}_0)=(0,0.1)$, $(0.5,0.5)$, and $(1.05,0.01)$ respectively. }\label{f1}
\end{figure}

\begin{figure}[h]
{\includegraphics[scale=0.33]{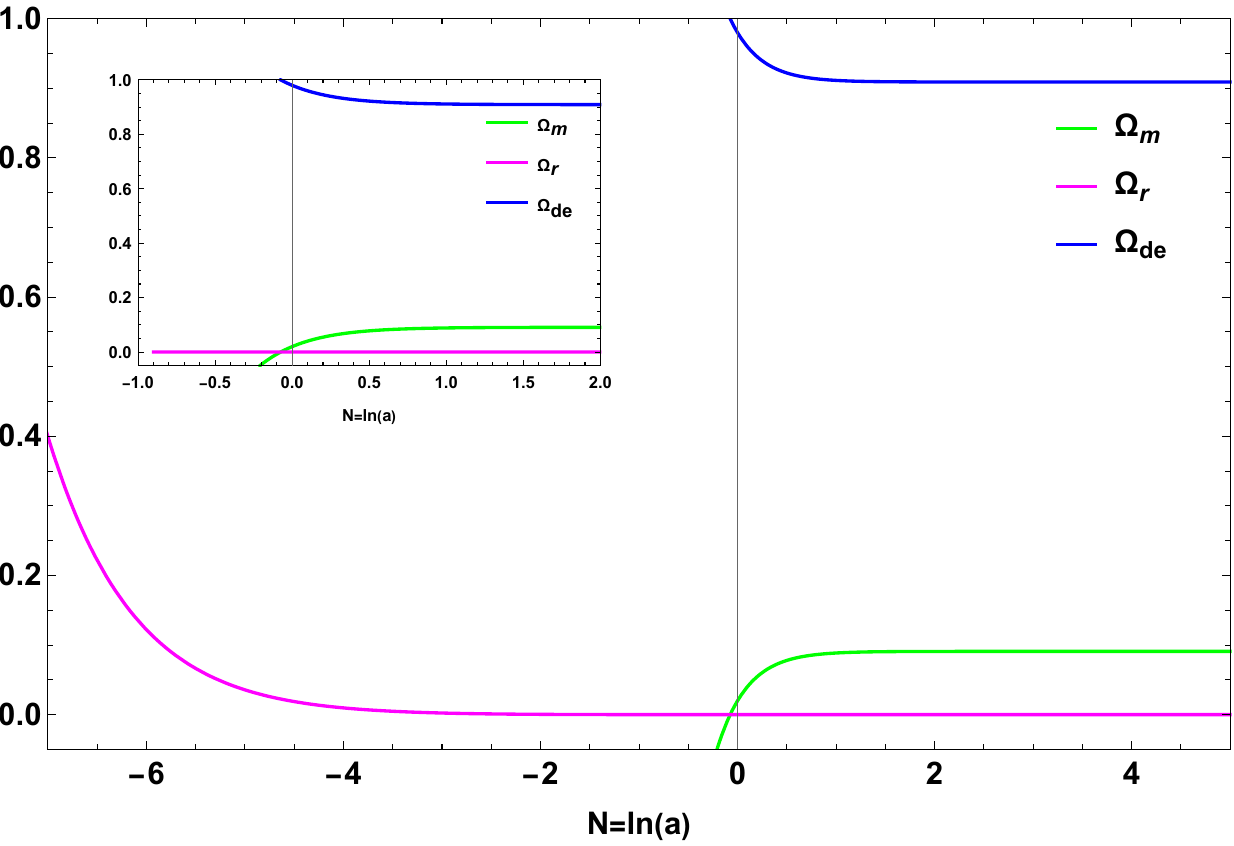}}
{\includegraphics[scale=0.35]{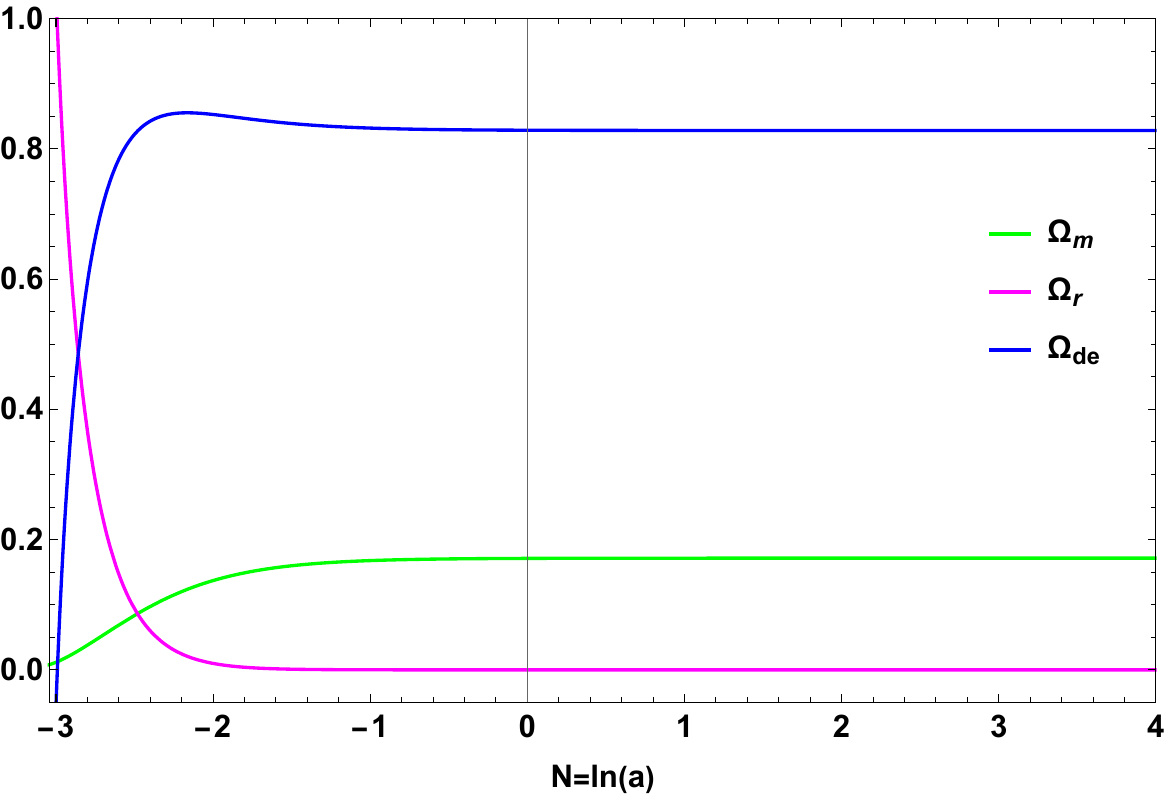}}
{\includegraphics[scale=0.34]{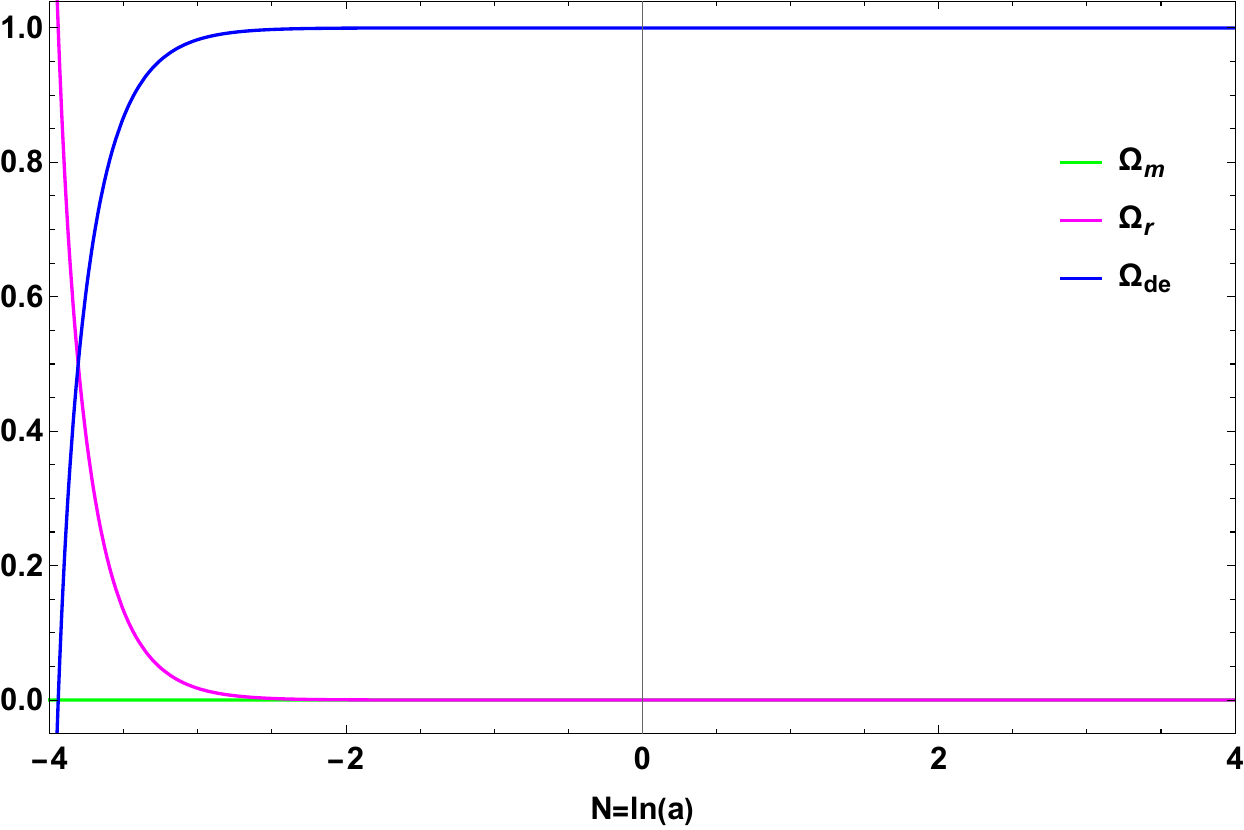}}
\caption{Profile of the matter, radiation, and the dark energy density parameter for the model I corresponding to cases $(s,\bar{\zeta}_0)=(0,0.1)$, $(0.5,0.5)$, and $(1.05,0.01)$ respectively.}\label{f2}
\end{figure}

\begin{figure}[H]
{\includegraphics[scale=0.4]{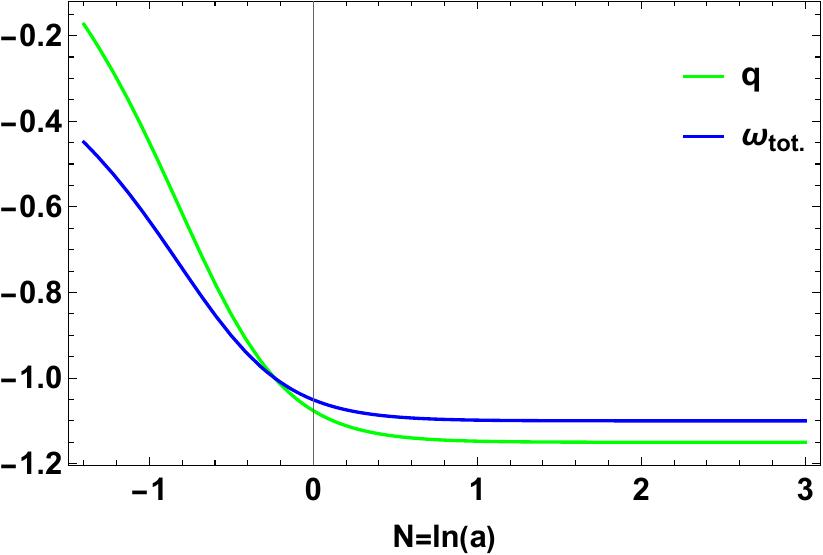}}
{\includegraphics[scale=0.38]{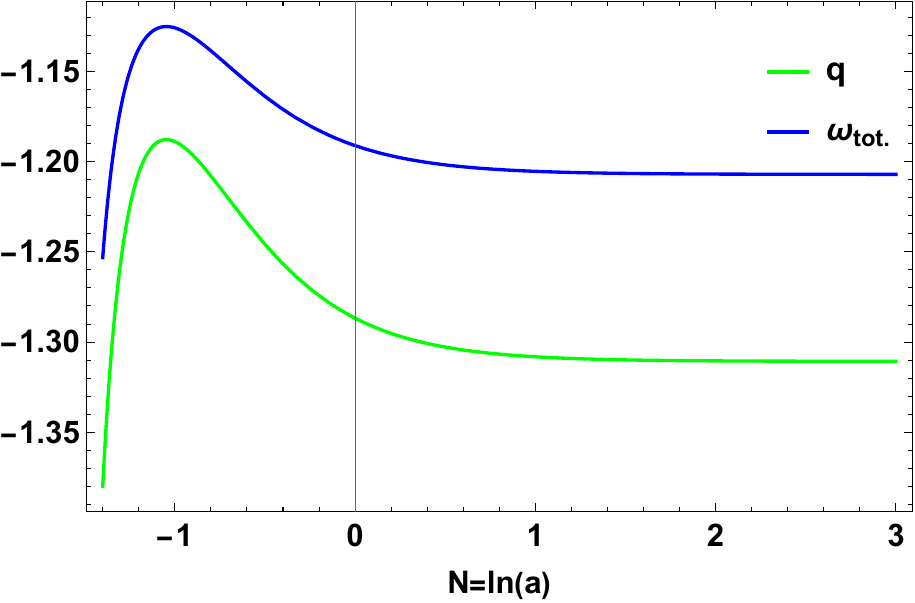}}
{\includegraphics[scale=0.45]{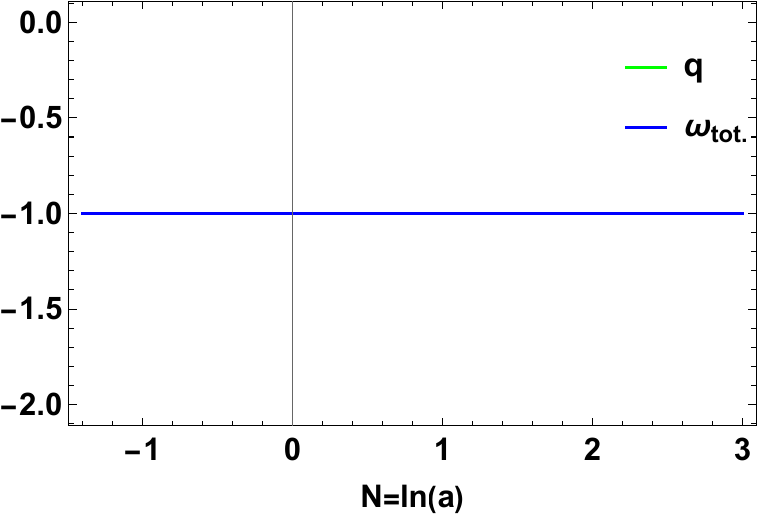}}
\caption{Profile of the effective equation of state and deceleration parameter for the model I corresponding to cases $(s,\bar{\zeta}_0)=(0,0.1)$, $(0.5,0.5)$, and $(1.05,0.01)$ respectively. }\label{f3}
\end{figure}

\justify \textbf{Model I: $f(Q)=-Q+\alpha Q^{-1}$ (i.e. $n=-1$) :} The considered model is characterized by value $n < 1$ (i.e., a correction to STEGR) that can provide modifications to the late-time cosmological phenomenon, potentially influencing the emergence of dark energy \cite{Jimenez2}. Corresponding to this model we investigate the presented dynamical system \eqref{4g}-\eqref{4h} for the aforementioned well-known cases of the exponent $s$, specifically $(s,\bar{\zeta}_0) = (0,0.1), (0.5,0.5)$ and $(1.05,0.01)$. For each of the cases, we linearize the system in the neighborhood of the obtained critical points and examine the corresponding stability. The outcome of the investigation corresponding to each case is presented in the following Table \eqref{Table-1} and the corresponding phase-space diagrams in Fig \eqref{f1}. Moreover, the behavior of the corresponding dimensionless density parameters, along with the effective equation of state and deceleration parameters, is presented in Fig \eqref{f2} and \eqref{f3}.

\begin{widetext}
\begin{table}[H]
\begin{center}\caption{Table shows the critical points and their behavior  corresponding to the model I.}
\begin{tabular}{|c|c|c|c|c|c|}
\hline
Cases $(s,\bar{\zeta}_0)$ & Critical Points $(x_c,y_c)$ & Eigenvalues $\lambda_1$ and $\lambda_2$ & Nature of critical point  & $q$ & $\omega_{total}$ \\
\hline 
$(0,0.1)$& $O_1(0.0909,0)$ & $ -4.3\:\: \text{and} \:\: -3.457$ & Stable  & $-1.15$ & $-1.1$ \\
\hline
$(0.5,0.5)$& $O_2(0.1715,0)$ & $-4.621 \:\: \text{and} \:\: -2.320$ & Stable  & $-1.00374$ & $-1.00249$ \\
\hline
$(1.05,0.01)$& $O_3(0,0)$ & $-4 \:\: \text{and} \:\: -3$ & Stable  & $-1$ & $-1$ \\
\hline
\end{tabular}\label{Table-1}
\end{center}
\end{table}
\end{widetext}

\textbf{Case I $(s=0,\bar{\zeta}_0=0.1)$} : Corresponding to the case I, we obtained single critical point $O_1(0.0909,0)$ with the associated eigenvalues $(\lambda_1,\lambda_2)=(-4.3,-3.457)$ and $(q,\omega_{total})=(-1.15,-1.1)$. Hence $O_1$ is stable equilibrium point representing the phantom type accelerated expansion of the universe with no transition epoch (see left panel of the Fig \eqref{f1}). The same is reflected in the evolutionary description of cosmological parameters shown in Fig \eqref{f2} and \eqref{f3} (left panel).

\textbf{Case II $(s=0.5,\bar{\zeta}_0=0.5)$} : Corresponding to the case II, we obtained single critical point $O_2(0.1715,0)$ with the associated eigenvalues $(\lambda_1,\lambda_2)=(-4.621,-2.320)$ and $(q,\omega_{total})=(-1.00374,-1.00249)$. Hence $O_2$ is stable equilibrium point representing the phantom type accelerated expansion of the universe with no transition epoch (see middle panel of the Fig \eqref{f1}). The same is reflected in the evolutionary description of cosmological parameters shown in Fig \eqref{f2} (right panel above) and \eqref{f3} (middle panel).

\textbf{Case III $(s=1.05,\bar{\zeta}_0=0.01)$} : Corresponding to the case III, we obtained single critical point $O_3(0,0)$ with the associated eigenvalues $(\lambda_1,\lambda_2)=(-4,-3)$ and $(q,\omega_{total})=(-1,-1)$. Hence $O_3$ is stable equilibrium point representing the de-Sitter type accelerated expansion of the universe with no transition epoch (see right panel of the Fig \eqref{f1}). The same is reflected in the evolutionary description of cosmological parameters shown in Fig \eqref{f2} (below) and \eqref{f3} (right panel).

\begin{figure}[H]
{\includegraphics[scale=0.41]{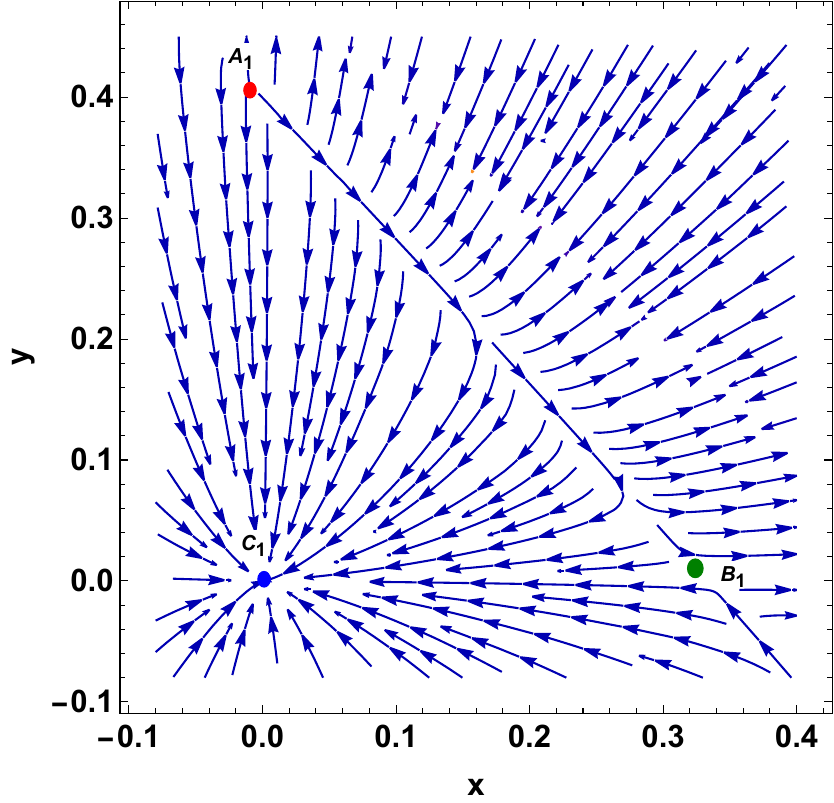}}
{\includegraphics[scale=0.315]{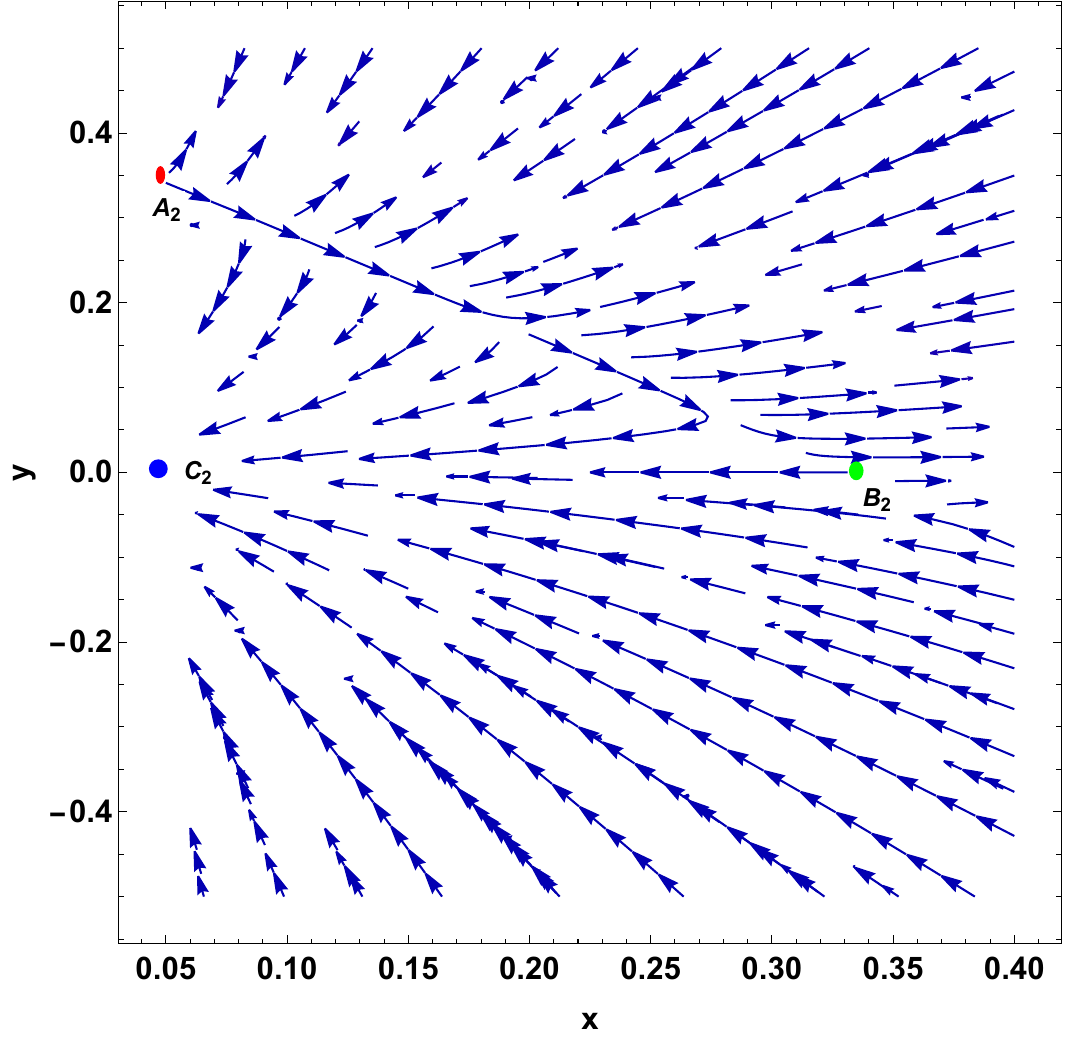}}
{\includegraphics[scale=0.42]{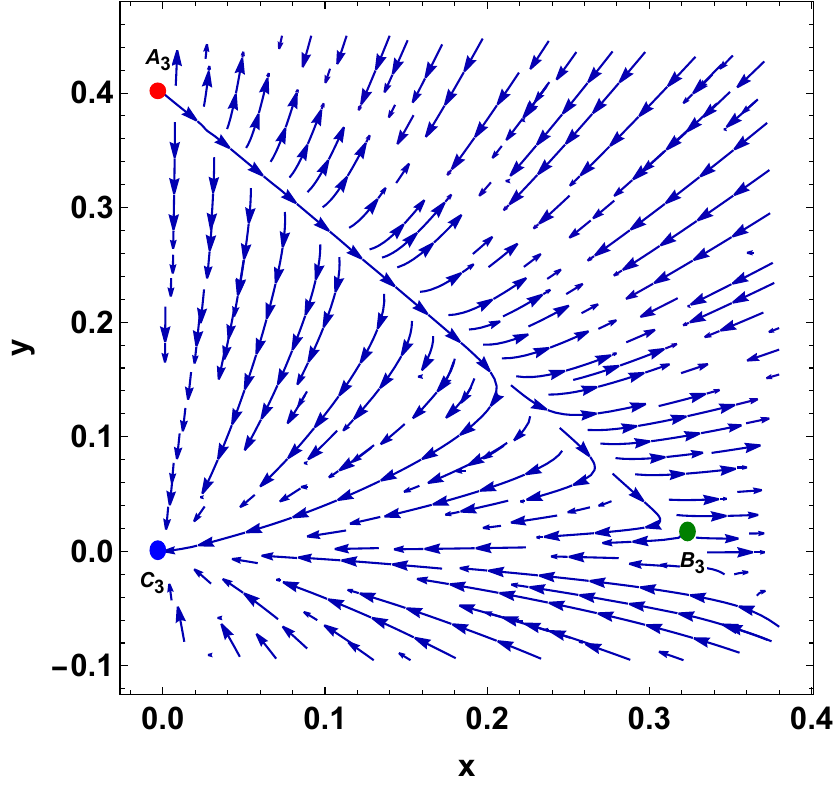}}
\caption{Phase-space diagrams representing the behavior of trajectories for the model II corresponding to cases $(s,\bar{\zeta}_0)=(0,0.001)$, $(0.5,0.005)$, and $(1.05,0.01)$ respectively.}\label{f4}
\end{figure}

\begin{figure}[H]
{\includegraphics[scale=0.48]{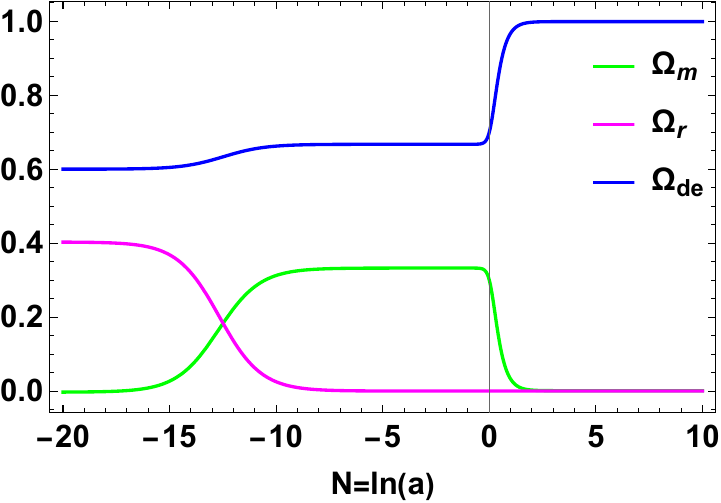}}
{\includegraphics[scale=0.48]{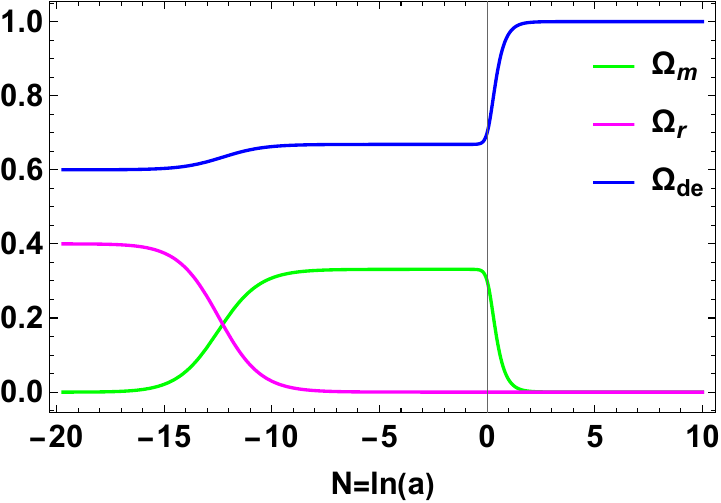}}
{\includegraphics[scale=0.485]{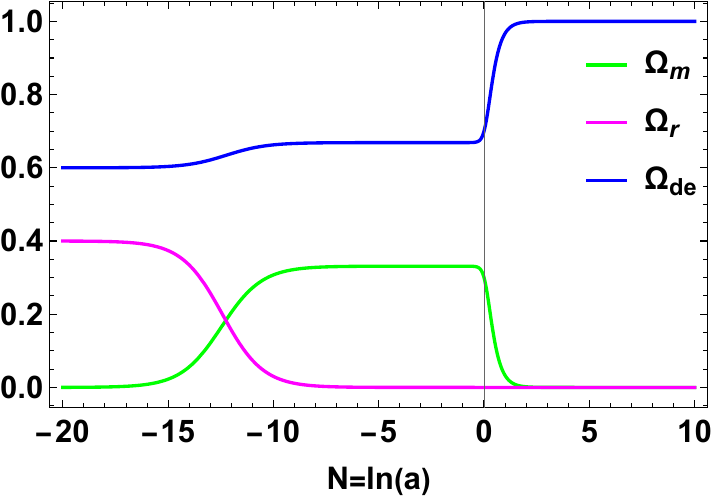}}
\caption{Profile of the matter, radiation, and the dark energy density parameter for the model II corresponding to cases $(s,\bar{\zeta}_0)=(0,0.001)$, $(0.5,0.005)$, and $(1.05,0.01)$ respectively.}\label{f5}
\end{figure}

\begin{figure}[H]
{\includegraphics[scale=0.35]{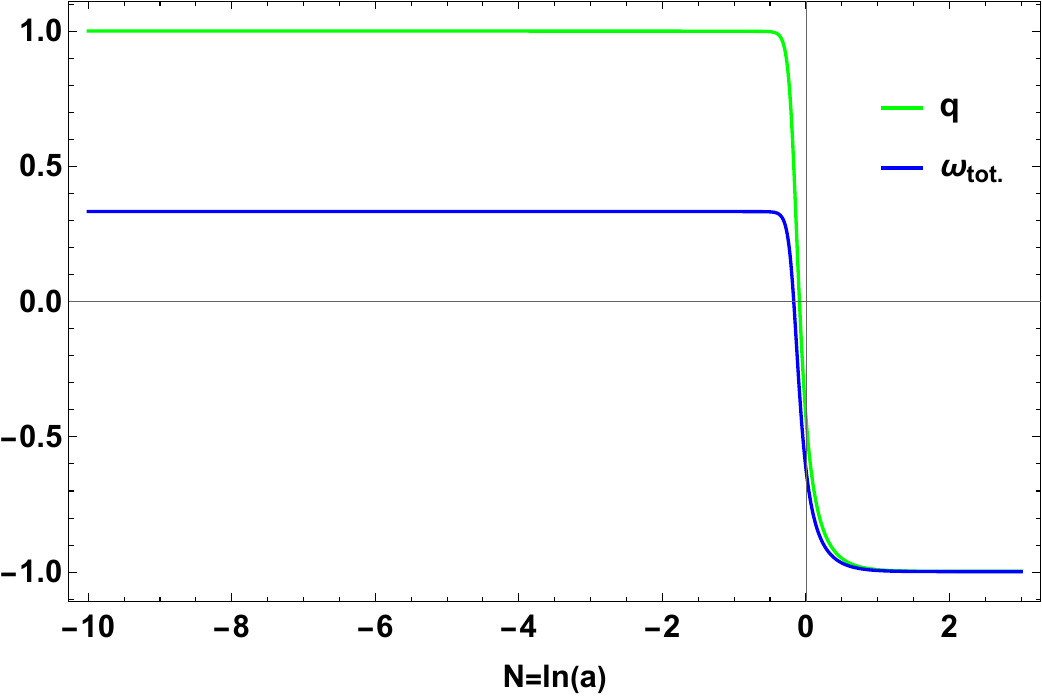}}
{\includegraphics[scale=0.35]{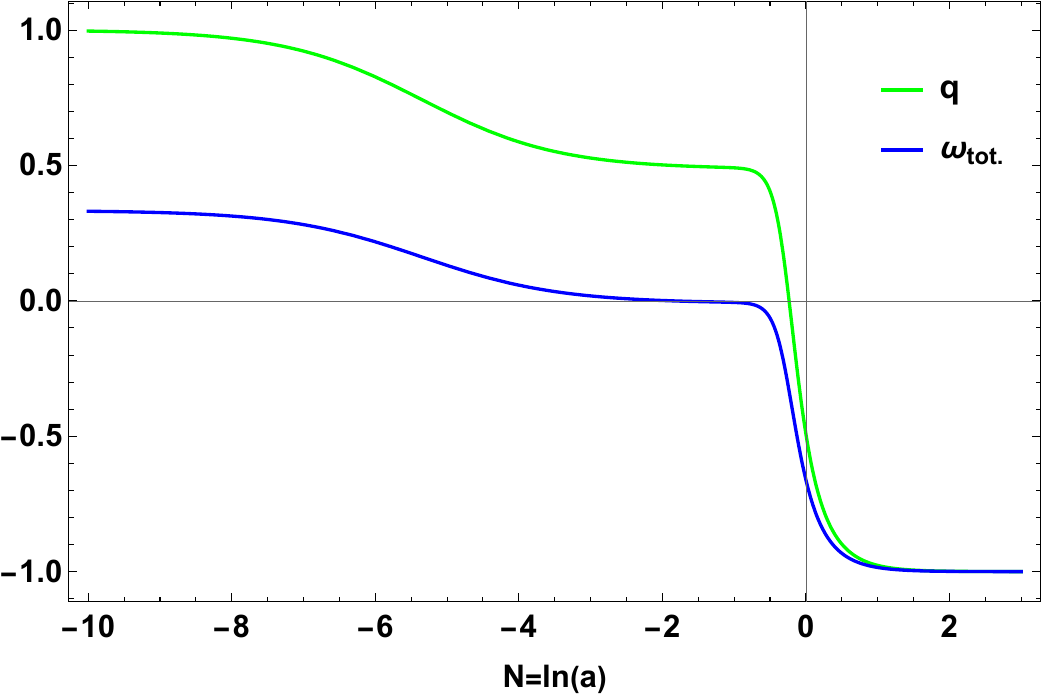}}
{\includegraphics[scale=0.35]{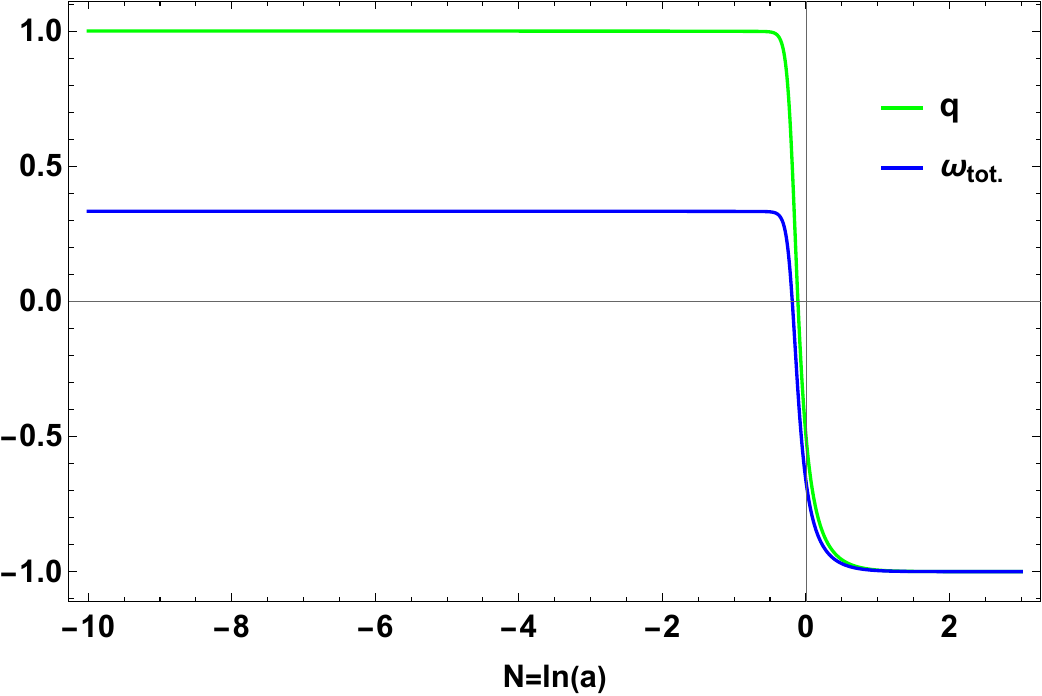}}
\caption{Profile of the effective equation of state and deceleration parameter for the model II corresponding to cases $(s,\bar{\zeta}_0)=(0,0.001)$, $(0.5,0.005)$, and $(1.05,0.01)$ respectively.}\label{f6}
\end{figure}

\justify \textbf{Model II: $f(Q)=-Q+\alpha Q^{2}$ (i.e. $n=2$) :} The considered model is characterized by value $n > 1$ (i.e., a correction to STEGR) that can provide modifications to the early universe phenomenon, offering potential inflationary solutions \cite{Jimenez2}. Corresponding to this model we investigate the presented dynamical system \eqref{4g}-\eqref{4h} for the aforementioned well-known cases of the exponent $s$, specifically $(s,\bar{\zeta}_0) = (0,0.001), (0.5,0.005)$ and $(1.05,0.01)$. The outcome of the investigation corresponding to each case is presented in the following Table \eqref{Table-2} and the corresponding phase-space diagrams in Fig \eqref{f4}. Moreover, the behavior of the corresponding dimensionless density parameters, along with the effective equation of state and deceleration parameters, is presented in Fig \eqref{f5} and \eqref{f6}.

\begin{widetext}
\begin{table}[H]
\begin{center}\caption{Table shows the critical points and their behavior corresponding to the model II.}
\begin{tabular}{|c|c|c|c|c|c|}
\hline
Cases $(s,\bar{\zeta}_0)$ & Critical Points $(x_c,y_c)$ & Eigenvalues $\lambda_1$ and $\lambda_2$ & Nature of critical point  & $q$ & $\omega_{total}$ \\
\hline 
& $A_1(0,0.403)$ & $ 19.976\:\: \text{and} \:\: 1.008$ & Unstable  & $1$ & $\frac{1}{3}$ \\
$(0,0.001)$& $B_1(0.332,0)$ & $ 8.919\:\: \text{and} \:\: -1.009$ & Saddle  & $0.495$ & $-0.003$ \\
& $C_1(0.001,0)$ & $ -3.993\:\: \text{and} \:\: -2.990$ & Stable  & $-0.996$ & $-0.997$ \\
\hline
& $A_2(0.02,0.4)$ & $19.85 \:\: \text{and} \:\: 1$ & Unstable  & $1$ &$ 0.3337$ \\
$(0.5,0.005)$ & $B_2(0.331,0)$ & $8.79 \:\: \text{and} \:\: -1.02$ & Saddle & $0.487$ & $-0.0086$ \\
& $C_2(0,0)$ & $-4 \:\: \text{and} \:\: -1.5$ & Stable & $-0.999$ & $-0.999$ \\
\hline
& $A_3(0,0.4)$ & $20 \:\: \text{and} \:\: 1$ & Unstable  & $1$ &$ \frac{1}{3}$ \\
$(1.05,0.01)$& $B_3(0.333,0)$ & $9.144 \:\: \text{and} \:\: -0.9716$ & Saddle  & $0.514$ & $0.00946$ \\
& $C_3(0,0)$ & $-4 \:\: \text{and} \:\: -3$ & Stable  & $-1$ & $-1$ \\
\hline
\end{tabular}\label{Table-2}
\end{center}
\end{table}
\end{widetext}

\textbf{Case I $(s=0,\bar{\zeta}_0=0.001)$} : Corresponding to the case I, the obtained critical points are $A_1(0,0.403)$,$B_1(0.332,0)$ and $C_1(0.001,0)$ with the associated eigenvalues $(\lambda_1,\lambda_2)=( 19.976,1.008)$,$( 8.919, -1.009)$ and $(-3.993,-2.990)$ respectively. Moreover, the associated $(q,\omega_{total})$ are $(1,\frac{1}{3})$, $(0.495,-0.003)$ and $(-0.996,-0.997)$ respectively. Hence the equilibrium point $A_1$ is unstable, $B_1$ is saddle and $C_1$ is stable, representing the radiation, matter, and de-Sitter phase respectively. It is evident from the direction of phase-space trajectories (see left panel of the Fig \eqref{f4}), the evolutionary trajectory of the model emerges out from the radiation epoch and then converges to the de-Sitter accelerated epoch via passing through the matter-dominated epoch i.e, having the evolution phase $A_1 \rightarrow B_1 \rightarrow C_1$. The evolutionary description of cosmological parameters shown in Fig \eqref{f5} and \eqref{f6} (left panel) reflects the same.

\textbf{Case II $(s=0.5,\bar{\zeta}_0=0.005)$}: Corresponding to the case II, the obtained critical points are $A_2(0.02,0.4)$, $B_2(0.331,0)$ and $C_2(0,0)$ with the associated eigenvalues $(\lambda_1,\lambda_2)=(19.85,1)$, $(8.79,-1.02)$ and $(-4,-1.5)$ respectively. Moreover, the associated $(q,\omega_{total})$ are $(1,0.337)$, $(0.487,-0.0086)$ and $(-0.999,-0.999)$ respectively. Hence the equilibrium point  $A_2$ is unstable, $B_2$ is saddle and $C_2$ is stable, representing the radiation, matter, and de-Sitter phase respectively. It is evident from the direction of phase-space trajectories (see middle panel of the Fig \eqref{f4}), the evolutionary trajectory of the model emerges out from the radiation epoch and then converges to the de-Sitter accelerated epoch via passing through the matter-dominated epoch i.e, having the evolution phase $A_2 \rightarrow B_2 \rightarrow C_2$. The evolutionary description of cosmological parameters shown in Fig \eqref{f5} and \eqref{f6} (middle panel) reflects the same.

\textbf{Case III $(s=1.05,\bar{\zeta}_0=0.01)$} :  Corresponding to the case III, the obtained critical points are $A_3(0,0.4)$,$B_3(0.333,0)$ and $C_3(0,0)$ with the associated eigenvalues $(\lambda_1,\lambda_2)=(20,1)$,$(9.144,-0.971)$ and $(-4,-3)$ respectively. Moreover, the associated $(q,\omega_{total})$ are $(1,\frac{1}{3})$, $(0.514,0.00946)$ and $(-1,-1)$ respectively. Hence the equilibrium point $A_3$ is unstable , $B_3$ is saddle and $C_3$ is stable, representing the radiation, matter, and de-Sitter phase respectively. It is evident from the direction of phase-space trajectories (see right panel of the Fig \eqref{f4}), the evolutionary trajectory of the model emerges out from the radiation epoch and then converges to the de-Sitter accelerated epoch via passing through the matter-dominated epoch i.e, having the evolution phase $A_3 \rightarrow B_3 \rightarrow C_3$.  The evolutionary description of cosmological parameters shown in Figures \eqref{f5} and \eqref{f6} (right panel) reflects the same.

\section{Conclusion}\label{sec5}
\justifying

The study of modified gravity incorporating non-metricity has garnered substantial attention in recent times, spanning various contexts such as black holes, wormholes, late-time observational constraints, and more. Meanwhile, viscous fluid cosmology has gained prominence for its explanation of the early stages of the universe and for addressing late-time acceleration. In this work, we have explored the function of the viscosity coefficient in the progression of cosmic evolution within the coincident $f(Q)$ gravity formalism. We begin with a non-linear function, specifically $f(Q)=-Q+ \Psi(Q) = -Q +\alpha Q^n$, where $\alpha$ and $n$ are arbitrary model parameters, and a newly proposed parameterization of the viscosity coefficient $\zeta$, specifically $ \zeta=\Bar{\zeta}_0 {\Omega^s_m} H $, where $\bar{\zeta}_0 = \frac{\zeta_0}{{\Omega^s_{m_0}}} $. The power-law model can describe the desired thermal history of the universe, both at the background and perturbation levels \cite{e11}. We obtained the set of autonomous differential equations by invoking the dimensionless density parameters as the governing phase-space variables corresponding to the assumed generic $f(Q)$ function. The proposed functional form involves a power-law correction to the STEGR scenario and holds considerable importance in both early and late-time cosmological contexts \cite{Jimenez2}. Specifically, the $f(Q)$ function under consideration with the value $n > 1$ can provide modifications to the early universe phenomenon, whereas the value $ n < 1 $ can provide modifications to the late-time cosmological phenomenon, potentially influencing the emergence of dark energy. Moreover, the assumed parameterization of the viscosity coefficient encompasses widely recognized models, $ \zeta=\zeta(H)$ for the case $ s=0$ and $ \zeta \sim {{\rho}_m}^{\frac{1}{2}} $ for the case $s = \frac{1}{2}$. Thus, for our analysis, we consider two toy models incorporating both corrections to the STEGR case with the aforementioned choices of the exponent $s$.

We consider the parameter choices for the model I (i.e. $f(Q)=-Q+\alpha Q^{-1}$ ) as $(s,\bar{\zeta}_0) = (0,0.1), (0.5,0.5)$ and $(1.05,0.01)$, whereas for the model II (i.e. $f(Q)=-Q+\alpha Q^2$ ) as $(s,\bar{\zeta}_0) = (0,0.001), (0.5,0.005)$ and $(1.05,0.01)$. The corresponding critical points and their behavior are presented in Table \eqref{Table-1} and \eqref{Table-2}, and the corresponding phase-space diagrams are presented in Fig \eqref{f1} and \eqref{f4}. Moreover, the evolutionary description of cosmological parameters such as density, deceleration, and the effective equation of state are presented in Fig \eqref{f2} and \eqref{f3} for the model I, and in Fig \eqref{f5} and \eqref{f6} for the model II. We found that corresponding to the model I, we obtained the stable de-Sitter type or stable phantom type (depending on the choice of exponent $s$) accelerated expansion of the universe with no transition epoch. Moreover,  the models with $n < 0$ have smaller value of $f_{\sigma8}$ in case of perfect fluid distribution, and hence, models with $n < 0$ are not favored by the data \cite{e11}. The same is reflected in our analysis as this case fails to describe cosmological epochs other than accelerated de-Sitter expansion. Further, for the model II, we obtained the evolutionary phase from the radiation epoch to the accelerated de-Sitter epoch via passing through the matter-dominated epoch. Hence, we conclude that model I provides a good description of the late-time cosmology but fails to describe the transition epoch, whereas model II modifies the description in the context of the early universe and provides a good description of the matter as well as radiation era along with the transition phase. Also, the analysis and outcomes of the present investigation favor the aforementioned results of the study \cite{Jimenez2} in the context of STEGR corrections. Moreover, one can reproduce the dynamics of $\Lambda$CDM by considering different $f(Q)$ function such as exponential case \cite{e12}. Furthermore, one can always find stable de-Sitter accelerated expansion for any choice of parameter $n$ $(n \neq 1)$ utilizing the non-trivial connections \cite{e13}.\\

\textbf{Data availability} There are no new data associated with this article.

\section*{Acknowledgments} 
 DSR acknowledges UGC, New Delhi, India for providing Junior Research Fellowship with (NTA-UGC-Ref.No.: 211610106591). RS acknowledges UGC, New Delhi, India for providing Senior Research Fellowship with (UGC-Ref. No.: 191620096030). PKS  acknowledges the Science and Engineering Research Board, Department of Science and Technology, Government of India for financial support to carry out Research Project No.: CRG/2022/001847. We are very much grateful to the honorable referees and to the editor for the
illuminating suggestions that have significantly improved our work in terms
of research quality, and presentation.



\begin{thebibliography}{90}

\bibitem{CANT} E. N. Saridakis et al., \textit{Modified Gravity and Cosmology: An Update by the CANTATA Network}, Springer (2021).

\bibitem{COSI} E. Abdalla et al., \textit{JHEAp} \textbf{34}, 49-211 (2022).

\bibitem{NEST} J. M. Nester and H. J. Yo, \textit{Chin. J. Phys.} \textbf{37}, 113 (1999).

\bibitem{JIM-1} J. B. Jim\'enez, L. Heisenberg, and T. Koivisto, \textit{Phys. Rev. D} \textbf{98}, 044048 (2018).

\bibitem{BARR} B. J. Barros, T. Barreiro, T. Koivisto, and N. J. Nunes, \textit{Phys. Dark Univ.} \textbf{30}, 100616 (2020).

\bibitem{ANAG} F. K. Anagnostopoulos, S. Basilakos, E. N. Saridakis, \textit{Phys. Lett. B} \textbf{822}, 136634 (2021).

\bibitem{NUNES} J. Ferreira, T. Barreiro, J. Mimoso, and N. J. Nunes, \textit{Phys. Rev. D} \textbf{105}, 123531 (2022).

\bibitem{NEOM} L. Atayde and N. Frusciante, \textit{Phys. Rev. D} \textbf{107}, 124048 (2023).

\bibitem{RODR}  J. T. S. S. Junior and M. E. Rodrigues, \textit{Eur. Phys. J. C} \textbf{83}, 475 (2023).

\bibitem{LAVI-1} F. D'Ambrosio, S. D. B. Fell, L. Heisenberg, and S. Kuhn, \textit{Phys. Rev. D} \textbf{105}, 024042 (2022).

\bibitem{SNEHA} S. Pradhan, S. Mandal, and P. K. Sahoo, \textit{Chin. Phys. C} \textbf{47}, 055103 (2023).

\bibitem{ZINNAT} Z. Hassan, S. Ghosh, P. K. Sahoo, and K. Bamba, \textit{Eur. Phys. J. C} \textbf{82}, 1116 (2022).

\bibitem{CAPE-1} F. Bajardi and S. Capozziello, \textit{Eur. Phys. J. C} \textbf{83}, 531 (2023).

\bibitem{PALIA-1}  N. Dimakis, A. Paliathanasis, and T. Christodoulakis, \textit{Class. Quantum Grav.} \textbf{38}, 225003 (2021).

\bibitem{ET} S. Sahlu and E. Tsegaye, \textit{arXiv}, arXiv:2206.02517 (2022).

\bibitem{ANAG-2}  F. K. Anagnostopoulos, V. Gakis, E. N. Saridakis, and S. Basilakos, \textit{Eur. Phys. J. C} \textbf{83}, 58 (2023).

\bibitem{ANDER} A. Lymperis, \textit{JCAP} \textbf{11}, 018 (2022).

\bibitem{CAPE-2} S. Capozziello and M. Shokri, \textit{Phys. Dark Univ.} \textbf{37}, 101113 (2022).

\bibitem{HOH} M. Hohmann, \textit{Phys. Rev. D} \textbf{104}, 124077 (2021).

\bibitem{WOM-2} W. Khyllep, A. Paliathanasis, and J. Dutta, \textit{Phys. Rev. D} \textbf{103}, 103521 (2021).

\bibitem{DE-1} G. Subramaniam, A. De, T. H. Loo, and Y. K. Goh, \textit{Fortschr. Phys.} \textbf{2023}, 2300038 (2023).

\bibitem{DE-2} G. Subramaniam, A. De, T. H. Loo, and Y. K. Goh, \textit{Phys. Dark Univ.} \textbf{41}, 101243 (2023).

\bibitem{PALIA-2} N. Dimakis, M. Roumeliotis, A. Paliathanasis, P. S. Apostolopoulos, and T. Christodoulakis,  \textit{Phys. Rev. D} \textbf{106}, 123516 (2022).

\bibitem{C.E.} C. Eckart, \textit{Phys. Rev.} \textbf{58}, 919(1940).

\bibitem{W.I.} W. Israel, J. M. Stewart, \textit{Phys. Lett. B} \textbf{58}, 213 (1976).

\bibitem{W.I.-2} W. Israel, \textit{Ann. Phys.} (N.Y.) \textbf{100}, 310 (1976).

\bibitem{W.I.-3} W. Israel, J. M. Stewart, \textit{Proc. R. Soc. Lond. B} \textbf{365}, 43 (1979).

\bibitem{IB-1} I. Brevik, \textit{Entropy} \textbf{2012(14)}, 2302-2310 (2012).

\bibitem{IB-2} I. Brevik and O. Gron, \textit{Astrophys. Space Sci.} \textbf{347}, 399 (2013).

\bibitem{IB-3} I. Brevik et al., \textit{Int. J. Mod. Phys. D} \textbf{26}, 1730024 (2017).

\bibitem{IB-4} I. Brevik, A. N. Makarenko, and A. V. Timoshkin, \textit{Int. J. Geom. Methods Mod.} \textbf{16}, 1950150 (2019)

\bibitem{IB-5} I. Brevik and B. D. Normann, \textit{Symmetry} \textbf{2020(12)}, 1085 (2020).

\bibitem{JM} N. D. J. Mohan, A. Sasidharan, and T. K. Mathew, \textit{Eur. Phys. J. C} \textbf{77}, 849 (2017).

\bibitem{AVS} A. V. Astashenok, S. D. Odintsov, and A. S. Tepliakov, \textit{Nucl. Phys. B} \textbf{974}, 115646 (2022).

\bibitem{MAT} A. Sasidharan and T. K. Mathew, \textit{Eur. Phys. J. C} \textbf{75}, 348 (2015).

\bibitem{COPE} E. J. Copeland, A. R. Liddle, and D. Wands, \textit{Phys. Rev. D} \textbf{57}, 4686-4690 (1998).

\bibitem{DE-3} H. Shabani, Avik De, T. H. Loo, \textit{arXiv}, arxiv:2304.02949 (2023).

\bibitem{WOM} W. Khyllep, J. Dutta, E. N. Saridakis, and K. Yesmakhanova, \textit{Phys. Rev. D} \textbf{107}, 044022 (2023).

\bibitem{Mishra-2} S. A. Kadam, B. Mishra, and J. L. Said, \textit{Eur. Phys. J. C} \textbf{82}, 680 (2022).

\bibitem{HAMID} H. Shabani and M. Farhoudi, \textit{Phys. Rev. D} \textbf{88}, 044048 (2013).

\bibitem{APLL} A. Paliathanasis, \textit{Phys. Dark Univ.} \textbf{41}, 101255 (2023). 

\bibitem{Jimenez3} J. B. Jimenez, L. Heisenberg, and T. S. Koivisto, \textit{Universe} \textbf{5}, 173 (2019).

\bibitem{KUHN} F. D'Ambrosio, L. Heisenberg, and S. Kuhn, \textit{Class. Quantum Grav.} \textbf{39}, 025013 (2021).

\bibitem{Jimenez2} J. B. Jimenez, L. Heisenberg, T. Sebastian Koivisto, and S. Pekar, \textit{Phys. Rev. D} \textbf{101}, 103507 (2020).

\bibitem{NVV} N. Cruz et al., \textit{arXiv}, arXiv:2304.12407 (2023).

\bibitem{e11} W. Khyllep et al., \textit{Phys. Rev. D} \textbf{107}, 044022 (2023).
 
\bibitem{e12} C. G. Boehmer, E. Jensko, and R. Lazkoz, \textit{arXiv}, arXiv:2303.04463v1 (2023).

\bibitem{e13} A. Paliathanasis, \textit{arXiv}, arXiv:2304.04219v2 (2023).



\end{thebibliography}
\end{document}